\documentclass[apjl]{emulateapj}
\usepackage{amsmath}

\newcommand{\Msolar}{M$_{\odot}$}

\begin{document}

\title{Interrupted Stellar Encounters in Star Clusters}
\shorttitle{Interrupted Stellar Encounters in Star Clusters}

\author{Aaron M.\ Geller$^{a,b}$}
\affil{Center for Interdisciplinary Exploration and Research in Astrophysics (CIERA) and Department of Physics and Astronomy, Northwestern University, 2145 Sheridan Rd, Evanston, IL 60208, USA}
\affil{Department of Astronomy and Astrophysics, University of Chicago, 5640 S. Ellis Avenue, Chicago, IL 60637, USA}
\email{$^a$a-geller@northwestern.edu}
\thanks{$^b$NSF Astronomy and Astrophysics Postdoctoral Fellow}

\author{Nathan W.\ C.\ Leigh$^{c}$}
\affil{Department of Astrophysics, American Museum of Natural History, Central Park West and 79th Street, New York, NY 10024}
\affil{Department of Physics, University of Alberta, CCIS 4-183, Edmonton, AB T6G 2E1, Canada}
\email{$^c$nleigh@amnh.org}

\shortauthors{Geller \& Leigh}

\begin{abstract}

Strong encounters between single stars and binaries play a pivotal role in the evolution of star clusters.
Such encounters can also dramatically modify the orbital parameters of binaries, exchange partners in and out of binaries, 
and are a primary contributor to the rate of physical stellar collisions in star clusters.  
Often, these encounters are studied under the approximation that they happen quickly enough and within a small enough volume
to be considered isolated from the rest of the cluster.  In this paper, we study the validity of this assumption through the analysis
of a large grid of single -- binary and binary -- binary scattering experiments.  For each encounter we evaluate the encounter 
duration, and compare this with the expected time until another single or binary star will join the encounter. 
We find that for lower-mass clusters, similar to typical open clusters in our Galaxy, the percent of encounters 
that will be ``interrupted'' by an interloping star or binary may be 20-40\% (or higher) in the core, though for typical globular 
clusters we expect $\lesssim$1\% of encounters to be interrupted.  Thus, the assumption
that strong encounters occur in relative isolation breaks down for certain clusters.  Instead, many strong encounters 
develop into more complex ``mini-clusters'', which must be accounted for in studying, for example, the internal dynamics of 
star clusters, and the physical stellar collision rate. 

\end{abstract}

\keywords{binaries: general --- galaxies: star clusters: general --- globular clusters: general --- open clusters and associations: general --- stars: kinematics and dynamics ---  methods: numerical}

\section{Introduction}\label{intro}

The evolution of (collisional) star clusters is often conceptualized, at a basic level, as being governed by the combination of the 
long-range cumulative effects of weak stellar encounters, known as ``two-body relaxation'', and the results of short-range 
strong stellar encounters between individual stars and binaries.  In Monte Carlo models for globular cluster (GC) evolution, this 
assumption is more than a conceptual convenience, and is inherent to the functionality of the code 
\citep[e.g.][]{spurzem:96,joshi:00,vasiliev:15}.  
Two-body relaxation allows stars to gradually exchange energy, 
which evolves the cluster towards thermal equilibrium, and drives the processes of evaporation, mass segregation and core collapse.  
It has long been known that close encounters with ``hard'' binaries (i.e., those with relatively large binding energy compared to 
the kinetic energies of cluster stars, \citealt{heggie:75}) can halt core collapse by donating energy to other stars in the encounter,
which can be given back to the cluster through two-body relaxation processes.  This type of strong encounter may
decrease the semi-major axis of the binary, and indeed, strong encounters can modify all orbital parameters of binaries, 
exchange stars into and out of binaries, and even result in physical stellar collisions.  Moreover, strong encounters are a key 
component to star cluster evolution \citep{hut:83a}, and they can alter a binary population from its characteristics at birth 
\citep[e.g.][]{ivanova:05,hurley:07,marks:11,geller:13a,geller:13b,geller:15,leigh:15}.

As such, the outcomes of single -- binary (1+2) and binary -- binary (2+2) encounters are well studied \citep[e.g.][]{heggie:75,hills:75,hut:83b,fregeau:04}, and 
more recently stellar encounters involving triples are also gaining importance \citep{leigh:12,leigh:13}.  
Apart from direct $N$-body star cluster simulations \citep{aarseth:03,wang:15},
it is typical to make the 
simplifying assumption that such encounters happen rapidly enough, and within a small enough volume, that they are
effectively isolated from the rest of the cluster.  With such assumptions, one can run many few-body scattering experiments,
each involving perhaps 3-6 stars, to derive statistical cross sections of the outcomes of such encounters \citep[e.g.][]{hut:83b,fregeau:04}, 
and apply this knowledge to help understand the more complex evolution of a full star cluster, which itself may contain
many hundreds to millions of stars.  

In this paper we investigate the validity of the assumption of treating these strong encounters as isolated. In reality 
the encounters occur within a star cluster, and most often in the dense cluster core, where they may not be allowed to 
progress fully on their own.  More specifically, we use the numerical scattering 
code \texttt{FEWBODY} \citep{fregeau:04}, to perform millions of isolated 1+2 and 2+2 scattering experiments, described in 
Section~\ref{method}.  We then compare the total encounter duration to the predicted time until another single or 
binary star would join, or ``interrupt'', this ongoing stellar encounter, as defined in Section~\ref{times}.  We find that for 
certain cluster parameters, the fraction of interrupted strong encounters can reach over 40\% (Section~\ref{frac}).  Finally in 
Sections~\ref{discuss}~and~\ref{conc} we discuss the implications of these findings and provide our conclusions.

\section{$N$-body Scattering Simulations} \label{method} 
 
We present results from $>10^7$ individual 1+2 and 2+2 numerical scattering experiments performed using the 
\texttt{FEWBODY} code \citep{fregeau:04}.  We choose parameters for these scattering experiments that are 
relevant for the cores of star clusters with total masses ($M_\text{cl}$) and half-mass radii ($R_\text{hm}$) covering the range 
of observed open clusters (OCs) and GCs in our Galaxy.  
Specifically, we sample a grid\footnote{\footnotesize
As is clear from the top panel of Figure~\ref{fig:fint}, Milky Way star clusters do not occupy this entire parameter grid, 
though the outlying grid points may be useful under other conditions.}
 in $M_\text{cl}$ extending from $10^2$ to $10^6$~\Msolar\ in steps 
of 0.5 in $\log_{10}(M_\text{cl}$~[\Msolar]~$)$, and in $R_\text{hm}$ from 1 to 10 pc in steps of 1 pc.  
We performed two sets of experiments over this grid, one drawing from a mass function appropriate for an old GC, 
at an age of 10 Gyr and $\text{[Fe/H]}=-1.5$, and the other drawing from a mass function more appropriate for an OC, 
at an age of 300 Myr and $\text{[Fe/H]}=0$.

For a single star in a given scattering experiment, we choose a stellar mass from a \citet{kroupa:93} initial mass 
function (IMF) between 0.1~\Msolar\ and the turnoff mass at the age of the cluster ($\sim$0.95~\Msolar\ at 10 Gyr and 
$\text{[Fe/H]}=-1.5$, and $\sim$2.96~\Msolar\ at 300 Myr and $\text{[Fe/H]}=0$).
For a binary, we first choose the primary mass ($m_1$) from the same mass function and within the same mass limits.
We then draw a mass ratio ($q=m_2/m_1$) from a uniform distribution to select the secondary mass ($m_2$),
and enforce the criteria that $q\leq1$ and $m_2>0.1$~\Msolar.  
We derive a radius for each star using this stellar mass and the cluster [Fe/H], following the method of \citet{tout:96},
which we provide to \texttt{FEWBODY} for identifying physical collisions during the encounters (see \citealt{fregeau:04} 
and \citealt{leigh:12} for more details).

We choose binary orbital elements from the observed distributions of binaries with solar-type primary stars in the 
Galactic field from \citet{raghavan:10}, which are also consistent with observations of solar-type binaries in OCs
\citep[e.g.,][]{geller:10,geller:13a,geller:12}.  Specifically, we draw orbital periods from a log-normal distribution 
(with a mean of $\log(P$~[days])~=~5.03 and $\sigma$~=~2.28), with a short period limit at the Roche radius \citep{eggleton:83} 
and a long-period limit at the ``hard-soft boundary'' of the core.  Thus we require detached hard binaries initially,
and assume that soft binaries are disrupted promptly and are generally not as relevant for the dynamical evolution of the cluster.
We estimate the maximum period for a given hard binary, $P_\text{hs}$, using the virial theorem, such that,
\begin{equation} \label{eq:phs}
P_\text{hs} = \frac{\pi G}{\sqrt{2}} \left(\frac{m_1 m_2}{m_3}\right)^{3/2} \left(m_1+m_2\right)^{-1/2}\sigma_0^{-3} ,
\end{equation}
where $m_3$ is the mass of the incoming object (either a single star or the combined mass of a binary), 
and $\sigma_0$ is the three-dimensional velocity dispersion (assumed to be $\sigma_0=\sqrt{3}\sigma_\text{0,1D}$). 
As we discuss in Section~\ref{discuss}, we ran additional experiments drawing binary orbital periods from a uniform distribution in $\log(P)$,
a common theoretical assumption.
Finally, we draw eccentricities from a uniform distribution, and choose all angles of the encounter randomly.  

For the majority of our experiments, we choose the impact parameter for a given encounter randomly from a uniform distribution 
between 0 and 1 times the binary semi-major axis in a 1+2 encounter, or the sum of the two binary semi-major axes for a 2+2 encounter.
We discuss additional experiments with larger impact parameters in Section~\ref{discuss}. 
The velocity at infinity for the incoming object is drawn from a lowered Maxwellian distribution (typical for 
star clusters), which is defined by the velocity dispersion and the escape velocity, both calculated 
at the center of a \citet{plummer:11} model (with the given $M_\text{cl}$ and $R_\text{hm}$).

The above parameters define discrete 1+2 and 2+2 scattering experiments that are appropriate for the cores of typical Milky Way star clusters.
For each of our 90 clusters, we perform $10^4$ 1+2 and the equivalent number of 2+2 unique numerical scattering experiments
(at two different cluster ages and two different period distributions, as well as a subset at higher impact parameters).
We discuss results from these $>10^7$ scattering experiments in the following sections.  

\begin{figure}[!t]
\plotone{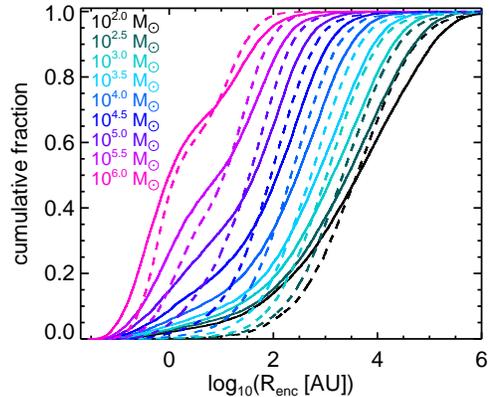}
\caption{
Cumulative distributions of the mean sizes ($R_\text{enc}$) of 1+2 (solid lines) and 2+2 (dashed lines) encounters  drawn from clusters 
with $R_\text{hm}=3$pc, over all $M_\text{cl}$ in our grid (as annotated in the figure), and using the 10 Gyr mass function.
\label{fig:rmean}
}
\end{figure}

\section{Encounter Timescales} \label{times}

There are two relevant timescales for the purposes of this study.  
The first is the encounter duration, which we will call $t_\text{d}$.  The second is the time until another single or binary star
will join, or ``interrupt'', this ongoing 1+2 or 2+2 encounter, which we will call $t_\text{e}$.
The first timescale, $t_\text{d}$, is calculated and outputted by \texttt{FEWBODY};
we refer the reader to \citet{fregeau:04},
and specifically Sections 3.3.4 and 3.3.5 and Equations 5 and 6,
for details.  In short, 
\texttt{FEWBODY} uses stability assessment techniques along with a few simple rules to determine when the 
separate components of the encounter (including stable hierarchies) will no longer interact with each other or evolve internally, at which 
point the encounter is deemed complete.
These criteria depend on a tolerance parameter $\delta = F_\text{tid}/F_\text{rel}$, the ratio of the tidal-to-binding (i.e., relative)
forces at apocenter,  which we set to the \texttt{FEWBODY} default value of $\delta=10^{-5}$.  
(In general, smaller values of $\delta$ yield more accurate results.)  
As we discuss in Section~\ref{discuss}, our results are only minimally sensitive to changes in $\delta$.

To calculate the second timescale, $t_\text{e}$, we must first find the time for a single ($_1$) or binary ($_2$) star to interrupt an ongoing
1+2 ($_3$) or 2+2 ($_4$) encounter, which we will refer to as, respectively, $\tau_{3+1}$,  $\tau_{3+2}$,  $\tau_{4+1}$,  and $\tau_{4+2}$.
We define these timescales as follows.
\begin{align} \label{eq:tau_enc1}
\beta &= \left(\frac{10^3\text{ pc}^{-3}}{n_0}\right) \left(\frac{v_\text{rms,0}}{5\text{ km s}^{-1}}\right) \left(\frac{0.5\text{ M}_{\odot}}{\langle m \rangle}\right) \left(\frac{1\text{ AU}}{R_\text{enc}}\right)\text{ ,} \nonumber \\
\tau_{3+1} &= 5.4 \times 10^{10} \left(1 - f_\text{b}\right)^{-1} \beta\text{ yr ,} \nonumber \\
\tau_{4+1} &= 4.2 \times 10^{10} \left(1 - f_\text{b}\right)^{-1} \beta\text{ yr ,} \nonumber \\
\tau_{3+2} &= 4.2 \times 10^{10} f_\text{b}^{-1} \beta\text{ yr ,} \nonumber \\
\tau_{4+2} &= 1.8 \times 10^{10} f_\text{b}^{-1} \beta\text{ yr ,} \nonumber \\
\end{align}
where $n_0$ is the central number density, $v_\text{rms,0}$ is the central root-mean-square velocity 
(and we assume $v_\text{rms,0}=\sigma_0=\sqrt{3}\sigma_\text{0,1D}$),
$\langle m \rangle$ is the mean mass of a star in the cluster, $R_\text{enc}$ is the mean radius of the encounter, (e.g., used to calculate the encounter's
mean geometric cross-section, $\pi$$R_\text{enc}^2$, see also Figure~\ref{fig:rmean}), 
and $f_\text{b}$ is the binary frequency. 
We derive these equations following the procedure in \citet{leonard:89}, with numerical pre-factors derived following \citet{leigh:11}.
Finally, the time until a subsequent encounter is given by,
\begin{equation}  \label{eq:tau_enc}
t_\text{e} = \begin{cases}
   \left(\Gamma_{3+1} + \Gamma_{3+2}\right)^{-1} & \text{, for 1+2} \\
   \left(\Gamma_{4+1} + \Gamma_{4+2}\right)^{-1} & \text{, for 2+2} ,\\
  \end{cases}
\end{equation}
where $\Gamma=1/\tau$.

Most of the parameters in Equation~\ref{eq:tau_enc1} come directly from the input values discussed above, and the assumption of 
a Plummer model for the cluster.  We describe our calculation of the remaining parameters below.

To calculate $n_0$ from the Plummer model, we first estimate the total number of stars 
by randomly drawing sufficient single stars\footnote{\footnotesize This method does not account for binaries, which may 
not have total masses drawn from the IMF, and therefore may slightly overestimate the true number density of objects.}
from the appropriate mass function (discussed above), to reach the cluster mass, $M_\text{cl}$.  
In order to reduce stochastic effects, we perform this estimate multiple times, until
the standard error on the mean number of stars derived for the cluster is $<0.01M_\text{cl}$.

We modified \texttt{FEWBODY} to calculate and output the mean of the maximum separation between any two 
stars in the encounter at every time step, and divide this value by two to get $R_\text{enc}$.  
At a given $R_\text{hm}$, the $R_\text{enc}$ distribution shifts toward lower values with larger $M_\text{cl}$ (see Figure~\ref{fig:rmean}). 
This is primarily due to the dependence of the hard-soft boundary on $M_\text{cl}$ (through $\sigma_0$), which shifts the distributions 
of binary semi-major axes, and also (by definition) the impact parameters,
towards smaller values at larger $M_\text{cl}$.

\begin{figure}[!t]
\plotone{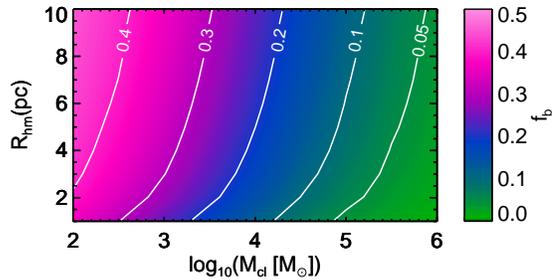}
\caption{
Contour plot of the binary frequencies ($f_\text{b}$) that result from our estimates of the hard-soft boundary, a field binary 
frequency of 50\%, and the 10 Gyr mass functions, for the given $R_\text{hm}$ and $M_\text{cl}$ values, as discussed in Section~\ref{method}.  
(The colors and lines show the same values.)
\label{fig:fb}
}
\end{figure}

Finally, to estimate the binary frequency, $f_\text{b}$, for a cluster with a given $M_\text{cl}$ and $R_\text{hm}$, we first 
estimate the expected average hard-soft boundary in the core of the cluster, as the mean of the
$P_\text{hs}$ values (from Equation~\ref{eq:phs}) for each encounter.
We then calculate the fraction of the log-normal input period distribution below this hard-soft boundary, $f_P$,  
and assume that, if the full period distribution could be occupied, this would result in a 50\% binary frequency \citep{raghavan:10}. 
To find the (hard) binary frequency for the cluster, we simply multiply $f_\text{b}=0.5f_P$.  In this way, 
we assume the total number of objects (binaries plus singles) in the cluster is fixed.  
The resulting $f_\text{b}$ values decrease toward higher $M_\text{cl}$ and lower $R_\text{hm}$ (Figure~\ref{fig:fb}), 
as is consistent with observed star clusters \citep[e.g.,][]{leigh:15b}.

\section{Fraction of Interrupted Stellar Encounters} \label{frac}

\begin{figure}[!t]
\plotone{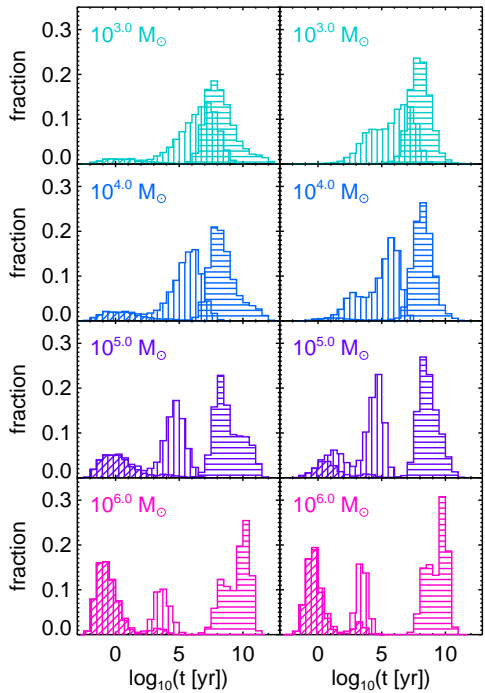}
\caption{
Distributions of the total encounter durations ($t_\text{d}$; vertically hatched histograms) and times until the next encounter 
($t_\text{e}$; horizontally hatched histograms) for the 1+2 (left panel) and 2+2 (right panel) scattering experiments drawn from 
clusters with $R_\text{hm}=3$pc and the $M_\text{cl}$ values indicated in each panel, for the 10 Gyr mass function.  
The diagonally hatched histograms show the encounters that resulted in at least 1 (left panel) or 2 (right panel) physical stellar collision(s).
All histograms are normalized by the total number of encounters. 
\label{fig:tdist}
}
\end{figure}

\begin{figure}[!t]
\plotone{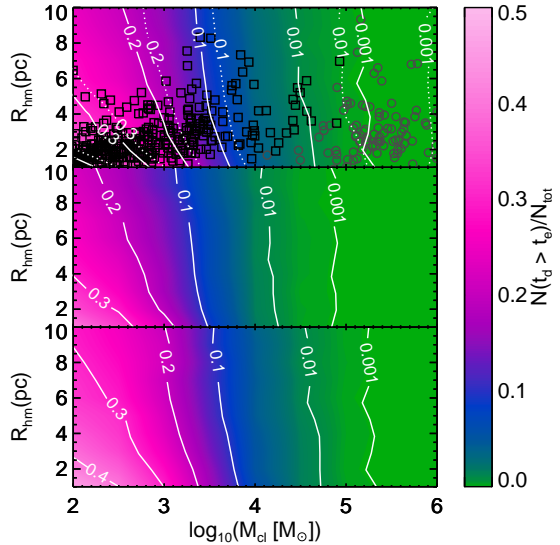}
\caption{
Contour plots showing the fraction of interrupted encounters
(where the encounter duration is greater than the predicted time until the next encounter: $t_\text{d}>t_\text{e}$, see Equation~\ref{eq:tau_enc})
for clusters with the given total masses, $M_\text{cl}$, and half-mass radii, $R_\text{hm}$.
The top panel shows the average fraction from all encounters, weighted by the 1+2 and 2+2 
encounter timescales from \citet{leigh:11}. 
The bottom and middle panels shows the separate results for our 1+2 and 2+2 scattering experiments, respectively.
The color shading and solid lines show (the same) results for the 10 Gyr (and sub-solar metallicity) 
mass function.  In the top panel we also show results for the 300 Myr (and solar metallicity) mass function with the dotted lines. 
For reference we also plot a sample of observed GCs \citep{harris:96} with the gray circles, and OCs \citep{piskunov:08,kharchenko:13} with the black squares.
(Note, for the OCs, we estimate $R_\text{hm}$ from the King core and tidal radii from \citealt{kharchenko:13}).
\label{fig:fint}
}
\end{figure}

We find from these scattering experiments that the percent of ongoing stellar encounters that are expected to be interrupted 
by an interloping single or binary star in a star cluster core ranges from below 1\% to over 40\%.    
The main results are plotted in Figures~\ref{fig:tdist}~and~\ref{fig:fint}.

For both the 1+2 and 2+2 experiments, we find that $t_\text{e}$ is only minimally sensitive to the
cluster mass (Figure~\ref{fig:tdist}).  Though the encounter's geometric cross-section ($\pi$$R_\text{enc}^2$), and binary frequency, both decrease with 
increasing cluster mass (see Figures~\ref{fig:rmean}~and~\ref{fig:fb}), the increase in central density roughly cancels out this effect.
On the other hand, the distribution of $t_\text{d}$ changes dramatically with cluster mass, shifting toward shorter 
encounter durations and developing a pronounced bimodal shape at higher $M_\text{cl}$.

The shift toward shorter encounter durations at higher $M_\text{cl}$ is due to the corresponding decrease in the semi-major axis at the 
hard-soft boundary, which results in tighter binaries involved in the encounters.  The encounter 
duration is then shorter due to the characteristically lower total angular momentum.

The emerging bimodal structure in the distribution of $t_\text{d}$ at higher $M_\text{cl}$ arises from an increasing frequency 
of physical collisions.  Indeed the fraction of encounters that result in a physical stellar collision in the cores of our 
most massive and centrally concentrated clusters reaches nearly 90\% (due to the more compact binaries involved in the encounters)\footnote{
\footnotesize 
Note that the collision frequency drops with larger impact parameters.}.
In the diagonally hatched histograms in Figure~\ref{fig:tdist}, we highlight the subset of encounters that end with $\leq2$ stars remaining
(due to collisions)
which clearly dominate the peaks at short $t_\text{d}$.  With only two (or in some cases one) stars remaining, an encounter is
considered to be finished, by definition,
and therefore on average will not last as long as a similar resonant encounter that does not result in a physical collision.

In Figure~\ref{fig:fint}, we compare $t_\text{d}$ and $t_\text{e}$ for each individual encounter directly (rather than as an ensemble).
In the core of a typical GC, at $M_\text{cl}\sim10^{5.5}$~\Msolar\ and $R_\text{hm}\sim3$~pc, $\lesssim$1\% of encounters are 
expected to be interrupted.  However, in the cores of typical OCs, 
at $M_\text{cl}\leq10^3$~\Msolar\ and $R_\text{hm}\leq3$~pc, the fraction of ongoing encounters that are expected to be 
interrupted by an incoming single or binary star is $\gtrsim25$\%, and reaches $>40$\% for the lowest mass and most compact 
clusters.

\section{Discussion} \label{discuss}

We find that a substantial fraction of strong encounters may be interrupted, while they are ongoing, by an interloping single or 
binary star in the cluster.  
In this case, the common assumption of encounters progressing fully in isolation breaks down, and instead the encounter develops into 
a ``mini-cluster'' with a more complicated dynamical evolution.
The fraction of interrupted encounters is largest for OCs 
(see Figure~\ref{fig:fint}), where, on average, the encounter geometric cross-sections are the largest (Figure~\ref{fig:rmean}) and the encounter 
durations are the longest (Figure~\ref{fig:tdist}).  For numerical models of star clusters, this result is encouraging, as
OCs are most often modeled with direct $N$-body methods \citep[e.g.,][]{aarseth:03}, 
which naturally account for stars or binaries interrupting an ongoing encounter.  
GCs are more often modeled with Monte Carlo codes \citep[e.g.,][]{chatterjee:10,hypki:13}, where stellar encounters are 
assumed to run to completion in isolation,
though direct $N$-body models are now approaching realistic GCs \citep{wang:15}.
Fortunately, as Figure~\ref{fig:fint} indicates, this assumption is much more valid 
(albeit not perfect) in massive GCs.  

In the following, we briefly discuss a few additional processes and parameters that may increase or decrease the fraction of interrupted encounters,
though all of which only contribute factors of approximately order unity. 

First, we did not account for any effects 
of two-body relaxation, which leads to mass segregation and the preferential loss of lower-mass single stars from the cluster,
and thereby tends to increase $\langle m \rangle$ and $f_\text{b}$ \citep[e.g.,][]{geller:13a}. 
Increasing $\langle m \rangle$ decreases $t_\text{e}$, and therefore tends to increase the fraction of interrupted encounters.
Increasing $f_\text{b}$ gives more weight to the 2+2 encounters in our calculations, and may increase or 
decrease the overall fraction of interrupted encounters accordingly.

Second, just as stars can be tidally stripped from a star cluster by the Galactic potential, 
stars involved in a strong encounter may endure an analogous process owing to the cluster potential.
Following this analogy, we estimate\footnote{\footnotesize Note that such simple analytic tidal radius estimates are 
not always reliable \citep{kupper:10,webb:13}; a rigorous investigation likely requires direct $N$-body star cluster simulations.} 
an effective tidal radius for each encounter, 
\begin{equation}\label{eq:rtid}
r_\text{t,eff}=\left(\frac{M_\text{enc}}{M_\text{core}}\right)^{1/3}r_\text{core}\text{ ,}
\end{equation}
where $M_\text{enc}$ is the total stellar mass of the encounter, $M_\text{core}$ is the core mass of the star cluster, $r_\text{core}$ is the 
core radius of the cluster, and we assume that the encounter occurs at $\lesssim 1 r_\text{core}$ from the cluster center. 
(Here, we estimate both $M_\text{core}$ and $r_\text{core}$ from the Plummer model.)  
If we assume that any encounter with, $R_\text{enc}>r_\text{t,eff}$ would be tidally disrupted before it is 
interrupted, this decreases the fraction of interrupted encounters by $\sim$15\% (almost independent of $M_\text{cl}$ and $R_\text{hm}$).  

We ran additional scattering experiments drawing binary orbital periods from a uniform distribution in $\log(P)$,
within the same period limits as for the log-normal period distribution.
For the same cluster parameters, this decreases the fraction of interrupted encounters by a factor of $\sim$2 
over nearly all $M_\text{cl}$ and $R_\text{hm}$, owing to the larger fraction of short-period binaries (with smaller geometric cross sections)
as compared to the more empirically motivated log-normal period distribution.

Additionally, we ran scattering experiments drawing impact parameters from uniform distributions extending 
from zero to 5, 10, 15 and 20 times
the size of the binary, respectively, and find nearly no change in the fraction of 
interrupted encounters from the results presented above.

We also ran scattering experiments with $\delta$ values of $10^{-7}$, $10^{-6}$, $10^{-5}$, $10^{-4}$, and $10^{-3}$, respectively.
We find that the fraction of interrupted encounters decreases slightly with smaller $\delta$ (i.e., more accurate outcome classifications 
in \texttt{FEWBODY}), and asymptotes at small $\delta$ toward a value of $\sim0.7$ times the results at our default $\delta=10^{-5}$.

In closing, one important implication of our results may be to increase the expected rate of physical stellar 
collisions between two (or more) stars in star clusters. 
In previous papers, we showed that, for fixed total energy and angular momentum, the probability of a physical collision during a stellar 
encounter increases as the number of stars in the encounter increases \citep{leigh:12,leigh:15}.
Thus if additional stars join an ongoing encounter, this should increase the probability of a collision. 
Moreover, collision rates (or cross sections) calculated from isolated scattering experiments may only be lower limits on the 
true collision rates in star clusters.

\section{Conclusions} \label{conc}

Strong stellar encounters in star clusters do not always run to completion without another stellar interloper cutting in on
the gravitational dance.
Indeed, we find that $\sim20-40$\% of encounters in the cores of OCs may be interrupted before completion (see Figure~\ref{fig:fint}).
Moreover, our results suggest that an assumption of isolated encounters is frequently invalid in the OC regime, and 
instead many encounters may develop into small-$N$ ``mini-clusters'', though this assumption may still be valid in more massive GCs.  
Finally, the probability of a physical stellar collision increases with an increasing number of stars in an encounter
\citep[e.g.][]{leigh:15}, and therefore these interrupted encounters may enhance the rate of stellar collisions in 
star clusters relative to what is assumed from isolated encounters alone.

\acknowledgments
A.M.G.\ is funded by a National Science Foundation Astronomy and Astrophysics Postdoctoral Fellowship under Award No.\ AST-1302765.
N.W.C.L.\ is grateful for the generous support of an NSERC Postdoctoral Fellowship.
This research was supported in part through the computational resources and staff contributions provided for the Quest high performance 
computing facility at Northwestern University which is jointly supported by the Office of the Provost, the Office for Research, and 
Northwestern University Information Technology.

\bibliographystyle{apj}

\end{document}